\documentclass{aa}

\usepackage{graphicx}
\usepackage{txfonts}
\usepackage{color}

\begin{document}

\title{The origin and properties of red spirals: Insights from cosmological simulations}

\author{Ewa L. {\L}okas
}

\institute{Nicolaus Copernicus Astronomical Center, Polish Academy of Sciences,
Bartycka 18, 00-716 Warsaw, Poland\\
\email{lokas@camk.edu.pl}}

\titlerunning{The origin and properties of red spirals}
\authorrunning{E. L. {\L}okas}


\abstract{
A significant fraction of spiral galaxies are red, gas-poor, and have low star formation rates (SFRs). We study these
unusual objects using the IllustrisTNG100 simulation. Among 1912 well-resolved disk galaxies selected from the last
simulation output, we identify 377 red objects and describe their properties and origins using a few representative
examples. The simulated red spirals turn out to be typically very gas-poor, have very low SFRs, are more metal-rich,
and have larger stellar masses than the remaining disks. Only about 13\% of red spirals suffered strong mass loss and
thus could have resulted from environmental quenching by ram pressure and tidal stripping of the gas, or similar
processes. The majority of red disks were probably quenched by feedback from the active galactic nucleus (AGN). This
conclusion is supported by the higher black hole masses and lower accretion rates of red disks, as well as the larger
total AGN feedback energies injected into the surrounding gas in the kinetic feedback mode implemented in the
IllustrisTNG simulations. The timescales of the gas loss correlate with the black hole growth for the AGN-quenched
galaxies and with the dark-matter loss for the environmentally quenched ones. The red spirals are more likely to
possess bars, and their bars are stronger than in the remaining disks, which is probably the effect of gas loss rather
than the reason for quenching.}


\maketitle

\section{Introduction}

The generally accepted classification system of galaxies, going back to \citet{Hubble1936}, divides these objects into
two broad classes: the flattened, star-forming, gas-rich, blue spiral galaxies and more spherical, quiescent, red
ellipticals. There is, however, a substantial number of spiral galaxies that do not form stars and are
predominantly red. The existence of such "anemic" spirals was first clearly pointed out by \citet{Bergh1976} and
illustrated with the prominent example of NGC 4921 in the Coma cluster. Numerous samples of such passive late-type
galaxies were later identified mostly among cluster populations using \textit{Hubble} Space Telescope observations
\citep{Couch1998, Poggianti1999, Wolf2009} and data from the Sloan Digital Sky Survey \citep[SDSS;][]{Goto2003}.

The first sample of red spirals that enabled a statistical analysis of these objects contained almost 300 galaxies and
was compiled within the Galaxy Zoo project \citep{Masters2010}. Nowadays, samples of red spirals include more than a
thousand galaxies, as was the case for those selected from the Galaxy and Mass Assembly (GAMA) survey
\citep{Mahajan2020} or those resulting from the combined analysis of the Hyper Suprime-Cam Subaru Strategic Program and
the GALEX-SDSS-WISE Legacy Catalog \citep{Shimakawa2022}. In this work we use the terms "red spirals" and "passive
spirals" interchangeably, although they are sometimes used to denote different objects. For example,
\citet{Cortese2012} demonstrated that some red disks are in fact forming stars at non-negligible rates and so cannot
really be considered passive.

The fraction of red objects among all spiral galaxies is surprisingly large and depends strongly on the stellar mass,
with more massive galaxies showing redder colors. However, the fractions differ depending on the sample: while
\citet{Masters2010} identified only about 5\% of red objects among the 5433 face-on spirals of the Galaxy Zoo, this
fraction was estimated to be as large as 20-50\% in the Two Micron All Sky Survey Extended Source Catalog
\citep{Bonne2015} and 42\% in the GAMA survey \citep{Mahajan2020}. In addition to their color, red spirals are
characterized by older stellar populations, less recent star formation, higher stellar masses, and lower dust
mass fractions than their blue counterparts \citep{Masters2010, Rowlands2012, Mahajan2020}.

Since the time of their discovery, red spirals have inspired a great deal of reflection as to their possible origin. Early on they
were found mostly in cluster environments, so it was natural to expect that a high-density neighborhood could contribute
to their loss of gas, the most obvious candidate being ram pressure stripping by the hot intracluster gas
\citep{Gunn1972}, including halo gas starvation \citep[when the galaxy gas supply is removed
from its halo rather than directly from the disk;][]{Bekki2002}, possibly combined with tidal stripping. Other external mechanisms
such as galaxy--galaxy interactions and mergers were also invoked. However, it was soon realized that the undisturbed
morphology of red spirals suggests that mechanisms other than strong interactions with the environment may be
responsible for their lack of gas and star formation.

Different studies performed so far have reached discrepant conclusions concerning the density of the medium most often
occupied by red spirals. While \citet{Mahajan2020} claim that the red spiral galaxies, especially those of lower mass, are
more likely to be found in high-density regions relative to their blue counterparts, \citet{Masters2010} found no clear
correlations between the properties of red spirals and environment, and \citet{Rowlands2012} discovered that the
passive spirals of the \textit{Herschel}-Astrophysical Terahertz
Large Area Survey (H-ATLAS) sample inhabit low-density environments similar to those of normal
spiral galaxies. The role of the environment in the formation of red spirals thus remains an open question.

Other possible explanations for the formation of passive spirals that rely on internal, rather than external,
factors suggest that they may just be old galaxies that used up all their gas during normal star formation activities,
which in its most extreme form is referred to as strangulation \citep{Peng2015}, or that they were quenched as a result
of feedback from supermassive black holes \citep{Okamoto2008}. In the later stages of active galactic nucleus (AGN)
activity, some of their energy is expected to heat the ambient gas, which suppresses star formation in the disk.

These processes were recently implemented in cosmological simulations of galaxy formation in the Universe performed
within the IllustrisTNG project \citep{Springel2018, Marinacci2018, Naiman2018, Nelson2018, Pillepich2018}. The
simulations follow the evolution of galaxies from early times to the present by solving gravity and hydrodynamics,
and applying, in addition to AGN feedback, prescriptions for star formation, galactic winds, and magnetic
fields. The set of simulations comprises the results obtained with different resolutions in boxes of size 300,
100, and 50 Mpc (labeled TNG300, TNG100, and TNG50, respectively).

The IllustrisTNG collaboration, building on the earlier experience from the Illustris project, improved the modeling of
AGN feedback, which led to a better agreement between the simulated and observed properties of the galaxy population,
including their morphology and colors \citep{Nelson2018, Genel2018, Rodriguez2019}. The new model adopted a two-mode
description of AGN feedback, with the thermal mode dominating at high accretion rates and the kinetic mode at low
accretion rates \citep{Weinberger2017, Weinberger2018}. The kinetic mode starts to be important when the black hole
mass exceeds about $10^8$ M$_\odot$; the feedback energy is then injected into the surrounding gas in a pulsed,
directed fashion rather than continuously, as is characteristic of the thermal mode. \citet{Weinberger2018}
demonstrated that the quenching of massive central galaxies in IllustrisTNG coincides with the onset of the kinetic
mode feedback.

The role of AGN feedback in the formation of red spirals was recently addressed by \citet{Xu2022}. They studied the
population of massive isolated disk galaxies in the largest simulation box, 300 Mpc per side, from the IllustrisTNG
set (TNG300) and concluded that kinetic feedback from AGNs is indeed responsible for quenching these galaxies. In this
paper we revisit the origin of red spirals using a higher-resolution simulation (TNG100) and selecting a sample
of red disks from all environments, without restricting the analysis to isolated objects. We thus aim to determine what
fraction of passive spirals were formed by environmental effects and what fraction by internal processes such as AGN
feedback. The higher resolution available in the 100 Mpc box allows us to include galaxies of lower stellar mass and
establish their morphology more reliably, as well as study their specific morphological features, such as bars. In Sect.
2 we describe our sample selection and the main properties of the identified red spirals. Section 3 is devoted to the
study of the origin of these galaxies in IllustrisTNG. In Sect. 4 we take a closer look at red disks with bars, and the
discussion follows in Sect. 5.

\section{Sample selection and basic properties}

For the purpose of this study, we used the publicly released data from the highest-resolution simulation performed in
the 100 Mpc box of the IllustrisTNG project (TNG100-1), as described by \citet{Nelson2019}. This simulation has
at redshift $z = 0$ the Plummer-equivalent gravitational softening length for the collisionless components (dark matter
and stars) $\epsilon = 0.74$ kpc and a median stellar particle mass of $1.6 \times 10^6$ M$_\odot$
\citep{Pillepich2018}. The TNG100-1 run provides a large collection of galaxies with sufficient resolution to
analyze samples of galaxies with different morphologies. Here we used the sample of 1912 disk galaxies selected from the
final ($z=0$) output of the IllustrisTNG100-1 simulation as described in \citet{Lokas2022}.

\begin{figure}
\centering
\includegraphics[width=8cm]{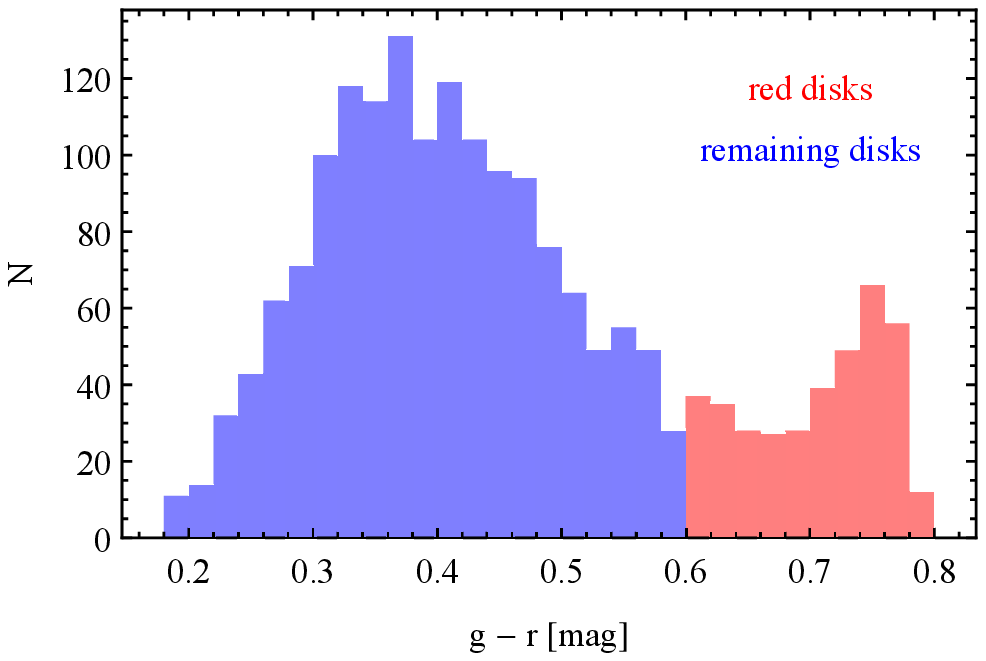}\\
\vspace{0.3cm}
\includegraphics[width=8cm]{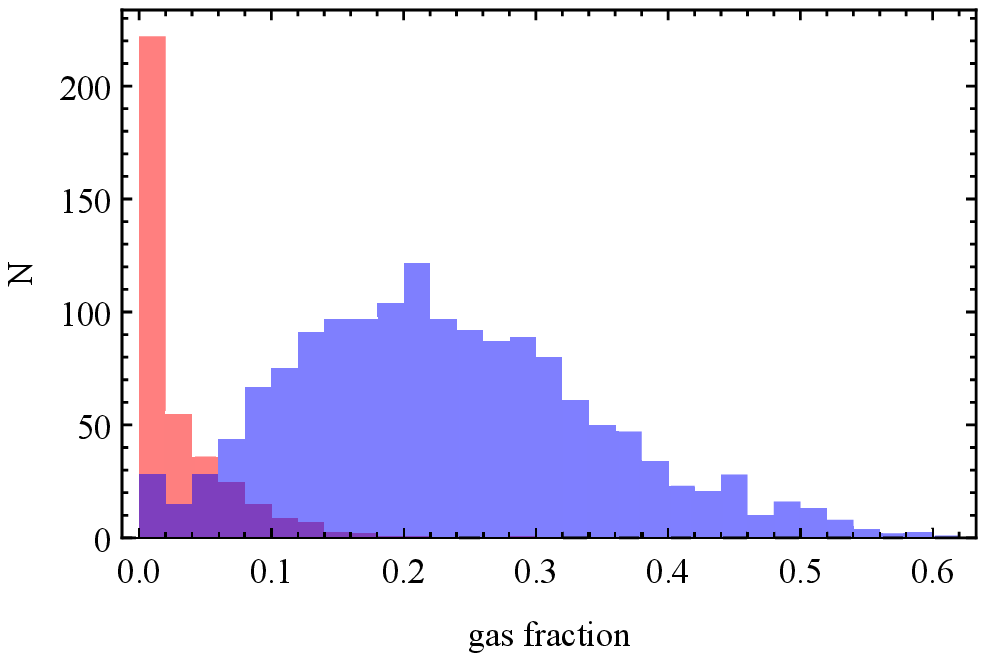}\\
\vspace{0.3cm}
\includegraphics[width=8cm]{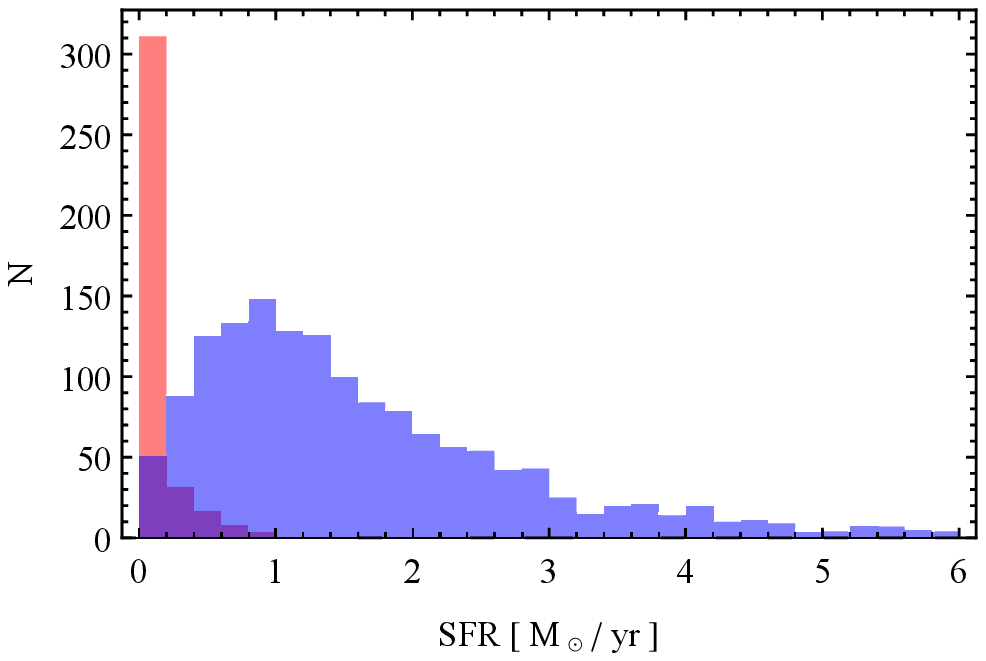}\\
\vspace{0.3cm}
\includegraphics[width=8cm]{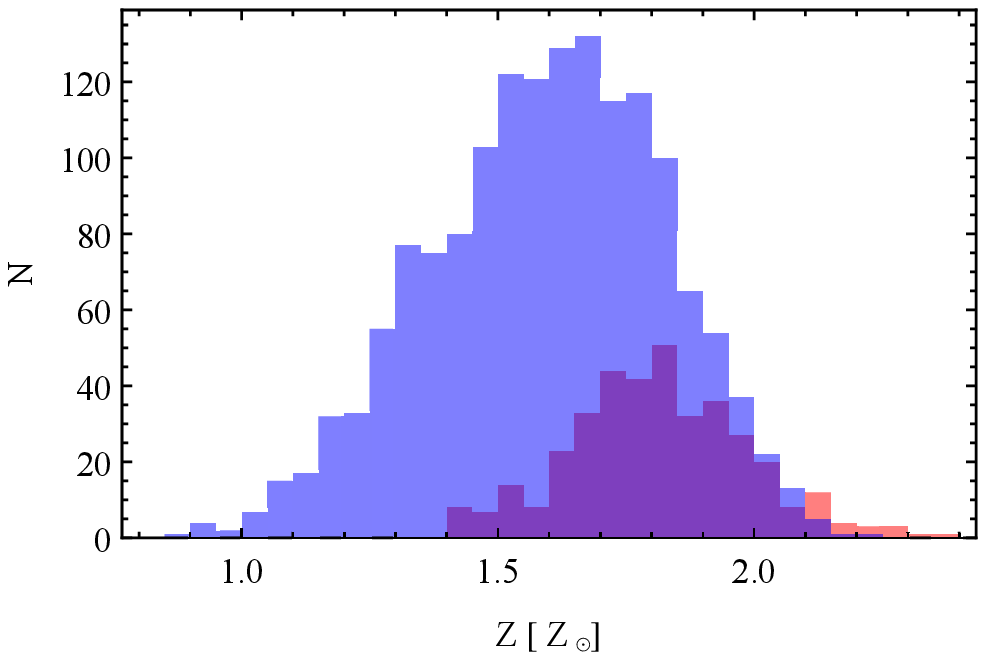}
\caption{Distributions of different properties of disky galaxies. The four panels, from top to bottom, show the
distributions of color, gas fraction, SFR, and metallicity for red disks with $g - r > 0.6$ (red) and
the remaining disks with $g - r < 0.6$ (blue). Measurements of the properties were done within $2 r_{1/2}$ except for the
color, which is estimated from all stars in the galaxy.}
\label{histogramssel}
\end{figure}

\begin{figure*}
\centering
\includegraphics[width=6cm]{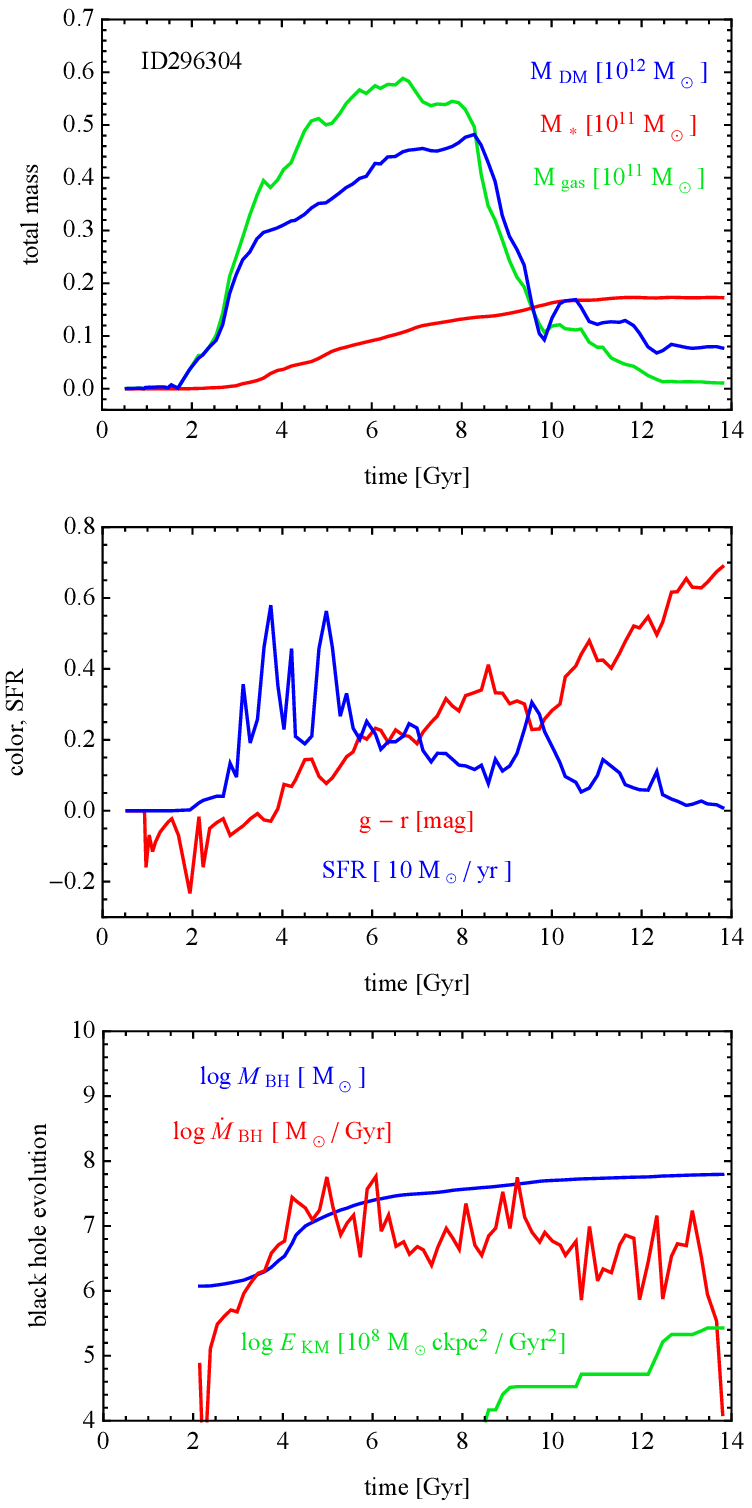}
\includegraphics[width=6cm]{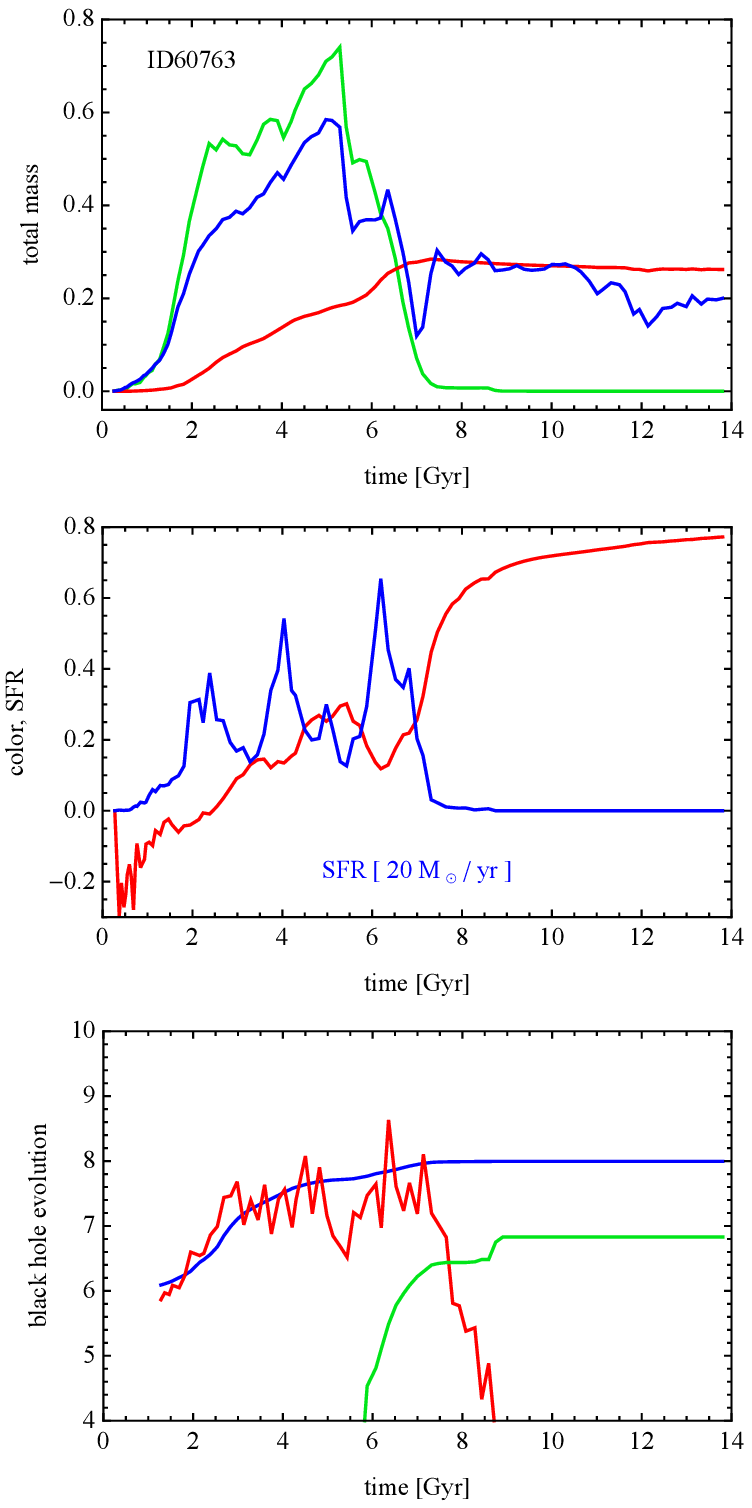}
\includegraphics[width=6cm]{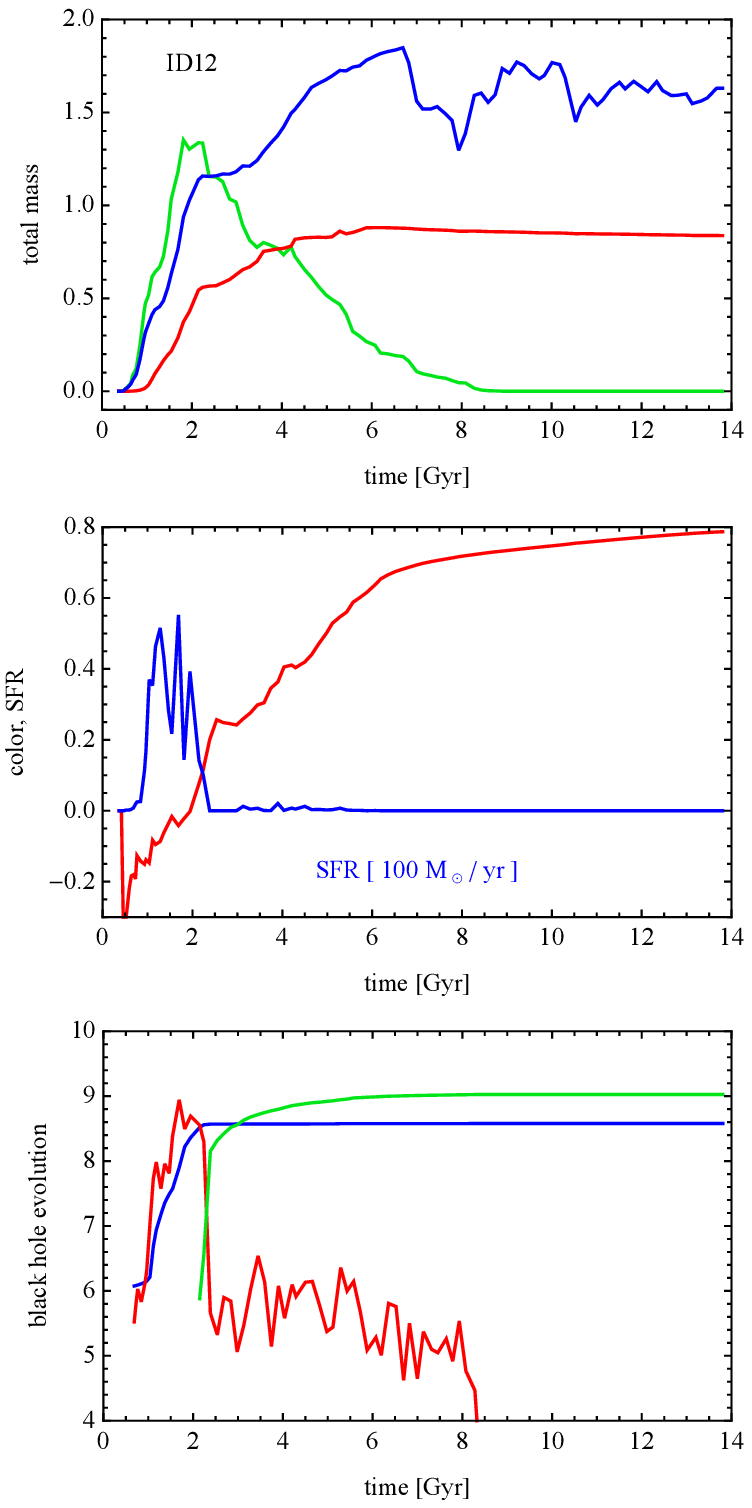}
\caption{Evolution of representative examples of red disk galaxies. The columns present the results for different
galaxies: ID296304, ID60763, and ID12. Upper row: Evolution of the total dark,
stellar, and gas mass shown with the blue, red, and green lines, respectively. Middle row: Evolution of color (red
line) and SFR (blue). Lower row: Evolution of the mass of the central supermassive black hole (blue), its accretion
rate (red), and the cumulative AGN feedback energy injected into
the surrounding gas in the kinetic mode (green). The units are as in the left column unless noted otherwise.}
\label{evolution}
\end{figure*}

Briefly, this selection was made by first choosing 6507 subhalos with total stellar masses greater than
$10^{10}$ M$_\odot$, which corresponds to about $10^4$ stellar particles per object. This criterion allows for the
reliable determination of the morphology of the galaxies. Next, the galaxies were assumed to be disky if they were
rotationally supported and rather thin. The sufficient amount of rotation was quantified, following \citet{Joshi2020},
by the rotation parameter $f > 0.4$, with $f$ defined as the fractional mass of all stars with circularity parameter
$\epsilon > 0.7$, where $\epsilon=J_z/J(E)$, $J_z$ is the specific angular momentum of the star along the angular
momentum of the galaxy, and $J(E)$ is the maximum angular momentum of the stellar particles at positions between 50
before and 50 after the particle in question in a list where the stellar particles are sorted by their binding energy.
The thickness of the disk was required to be $c/a < 0.5$, where $c/a$ is the shortest-to-longest axis ratio of the
stellar distribution within two stellar half-mass radii, $2 r_{1/2}$, estimated from the eigenvalues of the mass tensor
\citep{Genel2015}. We note that the thickness criterion excluded 433 galaxies from the 2345 galaxies selected with the
$f > 0.4$ criterion alone. While rather arbitrary, the $c/a < 0.5$ criterion was used to partially mitigate the problem
associated with the fact that the IllustrisTNG disks are too thick overall, with no galaxies having $c/a < 0.2$
\citep{Haslbauer2022}.

The red disks were selected from the total sample of 1912 disk galaxies using color estimates from
the catalogs of synthetic stellar photometry calculated including the effects of dust obscuration and provided with the
IllustrisTNG data \citep{Nelson2018}. We used the $g$ and $r$ colors from the SDSS $ugriz$ (rest-frame) bands and
required the galaxies to have $g - r > 0.6$, a threshold also adopted in other studies \citep{Nelson2018, Mahajan2020,
Lokas2020b} to separate red galaxies from blue ones. This color cut results in a sample of 377 red disks, which
we studied further. The distribution of this color measure in all the disk galaxies is shown in the upper panel of
Fig.~\ref{histogramssel}, divided into red ($g - r > 0.6$) and blue ($g - r < 0.6$) subsamples. The next two panels
present the distribution of other important properties that are expected to be different for the red and blue galaxies:
the gas fraction and the star formation rate (SFR). We note that while the color was estimated using all stars in the
galaxy, these two parameters were measured within two stellar half-mass radii because this is the region where star
formation typically occurs.

We can see that, as expected, most of the red galaxies do not contain any gas and do not form any stars within their
main body. However, there is a small tail in each of these distributions, with slightly higher values, though
much lower than the corresponding range of values for the blue galaxies. In particular, all the gas fractions for the
red disks are below 0.3, and only two are above 0.2. The present SFR values for red disks are all below 3.6 M$_\odot$
yr$^{-1}$, with only five galaxies having values above 1 M$_\odot$ yr$^{-1}$; for the remaining disks, they reach 17.8
M$_\odot$ yr$^{-1}$.

The last panel of Fig.~\ref{histogramssel} shows the distributions of metallicity. As we can see, they are not as
dramatically different as the previously discussed parameters. Both subsamples have smooth, Gaussian-like
distributions, but the red galaxies still show significantly higher metallicities on average, with a median of 1.8
$Z_\odot$ compared to the median for blue galaxies of 1.6 $Z_\odot$.

\begin{figure}
\centering
\includegraphics[width=8cm]{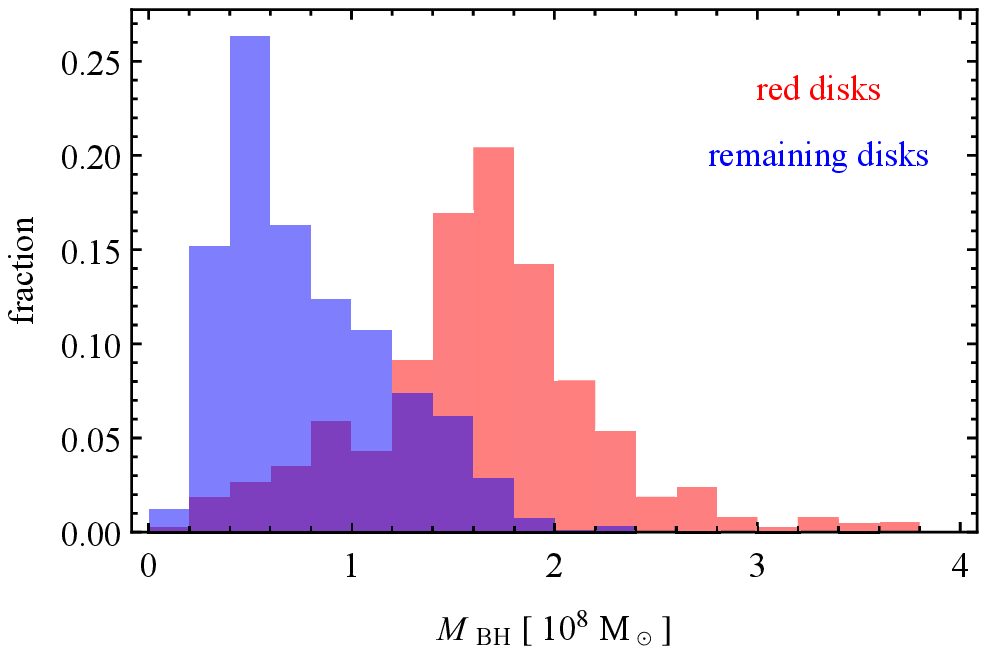}\\
\vspace{0.3cm}
\includegraphics[width=8cm]{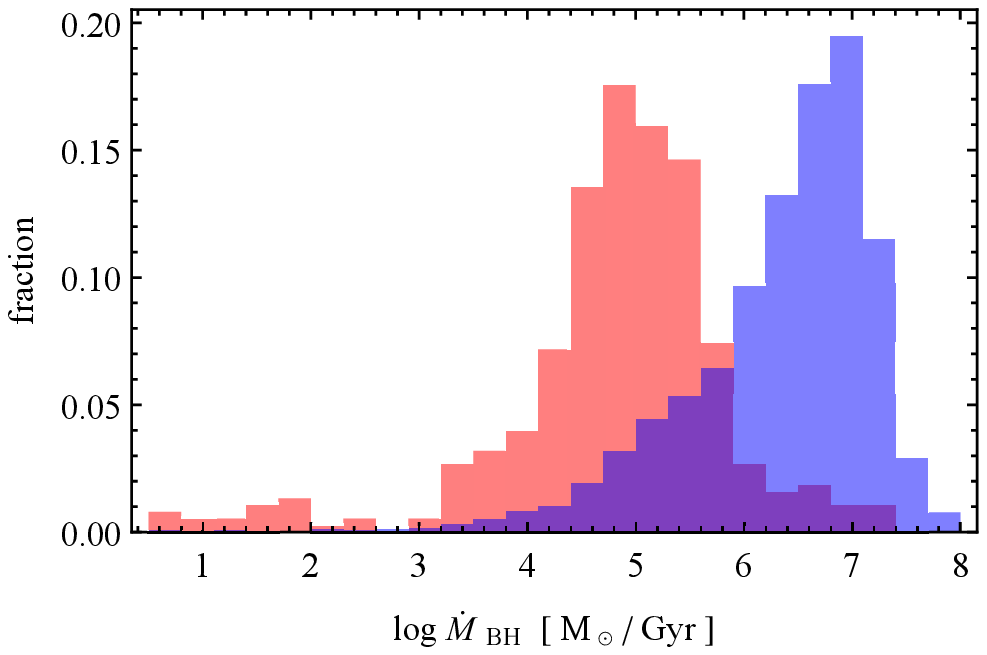}\\
\vspace{0.3cm}
\includegraphics[width=8cm]{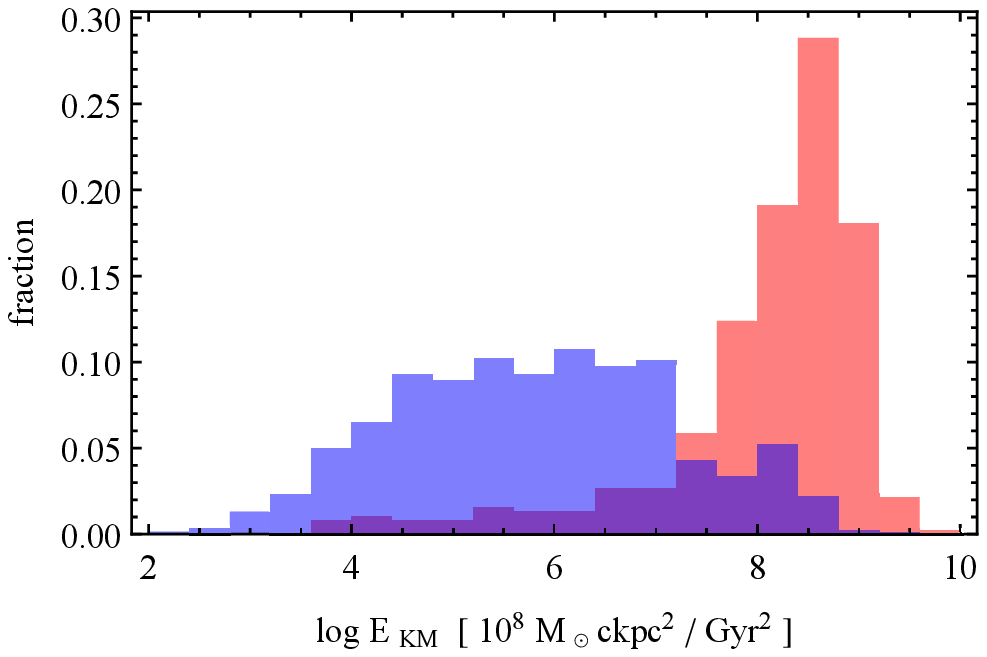}
\caption{Distributions of black hole properties of disky galaxies. The three panels, from top to bottom, show the
distributions of the black hole mass, the black hole accretion rate, and the total AGN feedback energy injected into
the surrounding gas in the kinetic mode for red disks with $g - r > 0.6$ (red) and remaining disks with $g - r < 0.6$
(blue). The histograms were normalized to unity.}
\label{histogramsbh}
\end{figure}

\section{Origin of red spirals}

Two main channels for the formation of red spirals have been proposed in the literature \citep{Masters2010,
Mahajan2020, Xu2022}. One is a scenario involving interactions, most likely with a more massive object. We
note that the rather regular disky morphology of red spirals excludes violent interactions such as recent major mergers
and close flybys. We expect that a spiral galaxy entering a halo of a bigger galaxy or a cluster will lose its gas due
to ram pressure and tidal stripping, which will soon cause it to cease its star formation and become red.
Another possibility of turning a star-forming galaxy into a red and dead one, especially applicable to isolated
objects, is related to the presence and activity of its supermassive black hole \citep{Xu2022}. The feedback from the
black hole is likely to expel most of the gas from the galaxy, which results in stopping the star formation. A
less interesting, but still viable, possibility is for the galaxy to just use up all the available gas and stop forming
stars.

Trying to distinguish between these scenarios for our sample of red disk galaxies in order to decide which galaxy
followed which path, we looked at the history of their total mass evolution and the evolution of their black hole mass,
accretion rate, and feedback energy. The interaction with a bigger object, likely to strip the galaxy of the gas, is
relatively easy to detect in the time dependence of its total dark-matter mass. Such an interaction leads to strong
mass loss as the dark mass in the outer part is assigned to the bigger halo. We note that at each simulation snapshot
the content of each subhalo is determined using the Subfind algorithm \citep{Springel2001}, which ensures that the
identified structures are self-bound. We only identified 48 galaxies, from our sample of 377 objects, that experienced
significant mass loss (they lost more than half of their maximum dark mass), which means that this is not a dominant
channel for the formation of red spirals.

A clear example of a galaxy for which this mechanism operated is the subhalo with identification number
ID296304 in the last simulation snapshot. The evolution of its different properties is shown in the left-column
plots of Fig.~\ref{evolution}. The changes in the dark mass in the upper panel indicate a strong interaction with a
more massive structure around $t = 10$ Gyr. Checking the neighborhood of the galaxy at that time, we found that
the galaxy did interact with a group of galaxies that contained the most massive object, with a mass on the order of
$10^{13}$ M$_\odot$. The time corresponds exactly to the pericenter passage of the galaxy on its orbit
around the group. As a result of the interaction, after an initial increase in the SFR (second panel), the galaxy gradually
lost its gas, stopped forming stars, and became red. Its black hole mass ($M_{\rm BH}$), however, remained well below
$10^8$ M$_\odot$ (lower panel), with a roughly constant accretion rate ($\dot{M}_{\rm BH}$), and the amount of AGN feedback
energy injected into the gas in the kinetic mode ($E_{\rm KM}$) was relatively low.

\begin{figure}
\centering
\includegraphics[width=7.5cm]{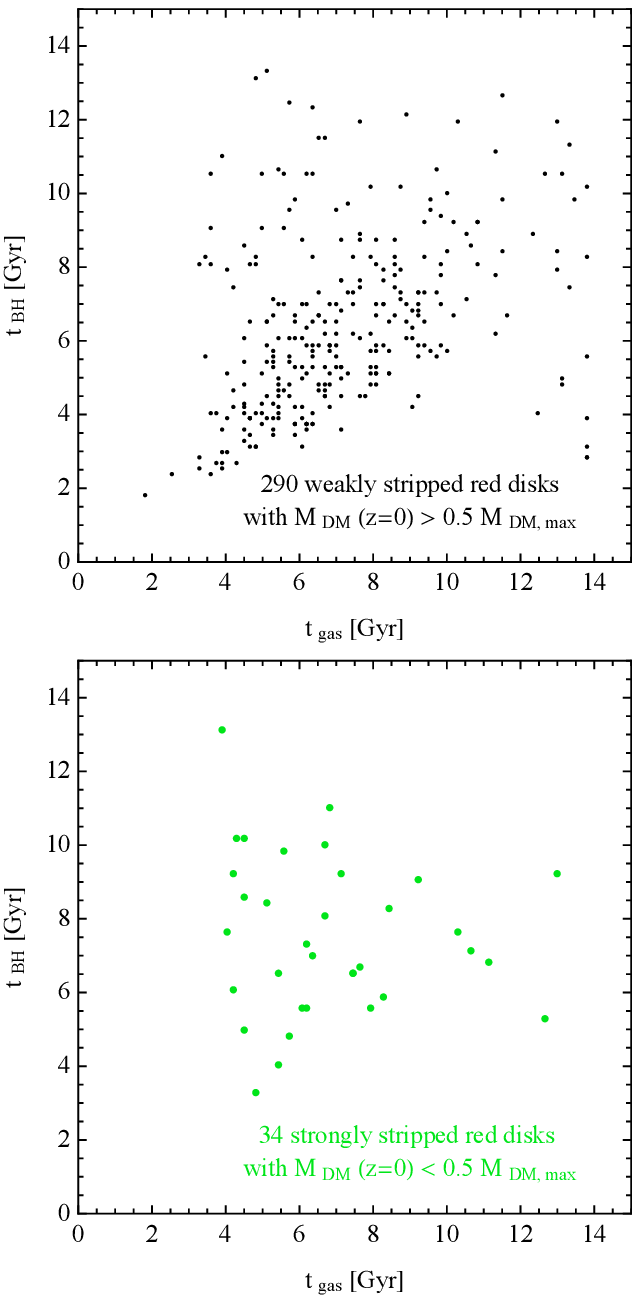}\\
\caption{Relation between characteristic timescales of evolution of gas and black hole in red disks. Upper
panel: Time when the black hole reaches the mass of $10^8$ M$_\odot$ versus the time of maximum gas mass for red disks
that lost less than half of their maximum dark mass. Lower panel: Same for red disks that lost more than half of their
maximum dark mass. Only 324 red disks with black hole masses larger than $10^8$ M$_\odot$ at present were included.}
\label{cortimegasbh}
\end{figure}

\begin{figure}
\centering
\includegraphics[width=7.5cm]{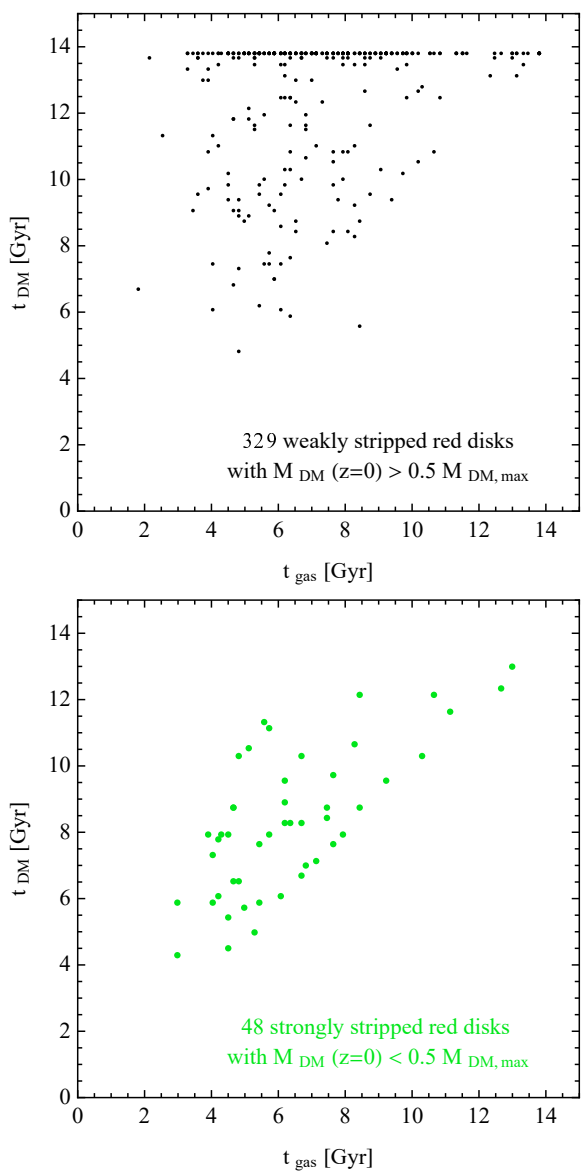}\\
\caption{Relation between characteristic timescales of evolution of gas and dark matter in red disks. Upper panel: Time
of maximum dark-matter mass versus the time of maximum gas mass for red disks that lost less than half of their maximum
dark mass. Lower panel: Same for red disks that lost more than half of their maximum dark mass. All 377 red disks were
included.}
\label{cortimegasdm}
\end{figure}

\begin{figure*}
\centering
\includegraphics[width=5.6cm]{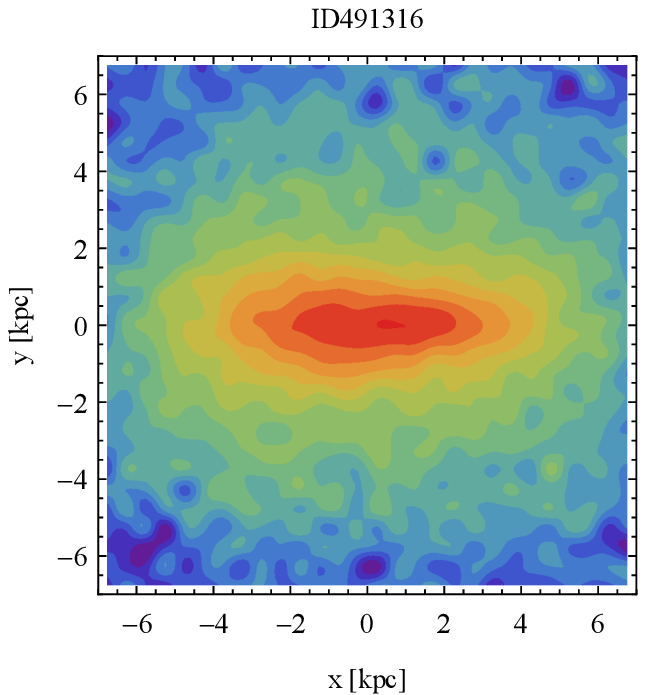}
\hspace{0.1cm}
\includegraphics[width=5.6cm]{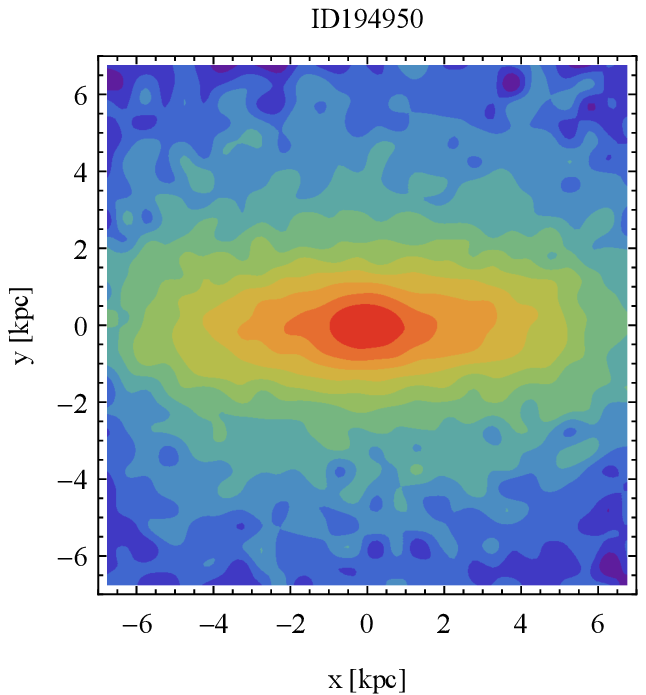}
\hspace{0.1cm}
\includegraphics[width=5.6cm]{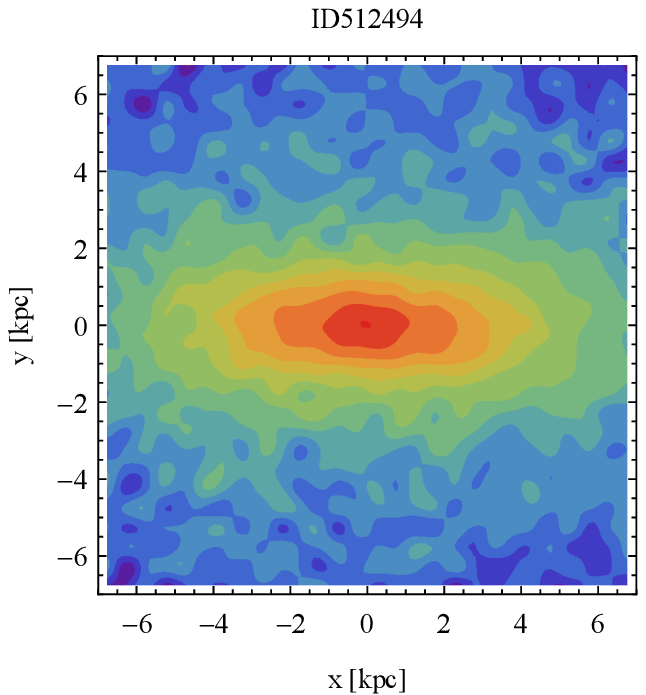}
\caption{Surface density distribution of stellar components of red disks with bars, ID491316, ID194950, and ID512494,
in the face-on view at the present time. The surface density, $\Sigma,$ is normalized to the central maximum value in
each case, and the contours are equally spaced in $\log \Sigma$.}
\label{surden}
\end{figure*}

\begin{figure}
\centering
\includegraphics[width=7.5cm]{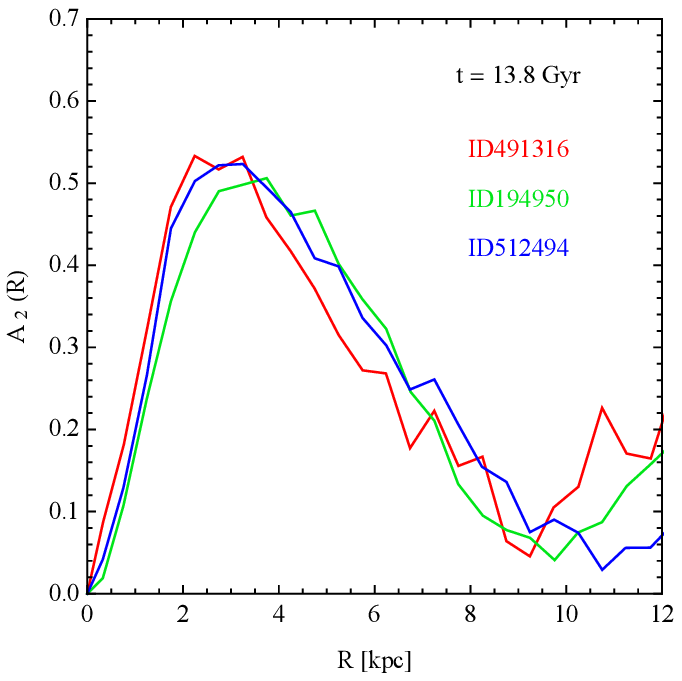}\\
\caption{Profiles of the bar mode, $A_2 (R)$, for selected red disks with bars: ID491316, ID194950, and ID512494.
Measurements were carried out for the present time in bins of $\Delta R = 0.5$ kpc.}
\label{a2profiles}
\end{figure}

The second channel for the formation of red spirals is well illustrated by the evolution of the properties of the
galaxy ID12 in the right column of Fig.~\ref{evolution}. The first panel of this set of plots shows that this galaxy
did not undergo any significant dark-matter  mass loss, meaning that it did not have any strong interactions with
massive neighbors. Instead, its supermassive black hole experienced very high accretion rates and strong growth early
on, between $t=1$ and 2 Gyr, as shown in the last panel of Fig.~\ref{evolution}. The black hole grew to be
more than $10^{8}$ M$_\odot$ over this short timescale, which resulted in a very large amount of AGN feedback energy
injected into the gas and the gradual expulsion of the gas. The SFR decreased and the galaxy became red (middle
panel).

These two clear examples are not intended to suggest that the sample of all red disks studied here can be
convincingly divided into two distinct subsamples using a similar analysis. Unfortunately, many cases are intermediate:
they show both mass loss and black hole feedback occurring at the same time. An example of such an object is
ID60763 (middle column of Fig.~\ref{evolution}). As indicated by the dark mass evolution in the upper panel, at around $t
= 7$ Gyr this galaxy interacted with a cluster-size object with a mass on the order of $10^{14}$ M$_\odot$ and a few other
massive galaxies. The evolution of the total gas mass shows that the gas was almost completely stripped around the
same time. This resulted in an almost immediate stopping of the star formation processes and the abrupt increase in the
$g-r$ color toward red values, as shown in the second panel of the column. The last panel of the column shows that
the loss of gas also resulted in a strong drop in the black hole accretion rate, but the black hole
mass almost reached the value of $10^{8}$ M$_\odot$ and the AGN feedback energy injected into the gas was significant.

Given that most of the galaxies in our red disk sample are isolated and did not experience significant interactions in
their history, it seems that the scenario involving the feedback from a supermassive black hole is the dominant  one.
This hypothesis is supported by a comparison between the distributions of the black hole mass, the accretion rate, and
the total AGN feedback energy injected into the surrounding gas in the kinetic mode for the red and remaining disks
shown in Fig.~\ref{histogramsbh}. We can see that the red disks typically have much higher black hole masses at present
than the remaining disks, with a median of $1.7 \times 10^{8}$ M$_\odot$ for the red sample and $0.7 \times 10^{8}$
M$_\odot$ for the rest of the disks. Only one galaxy in each of these subsamples of disks does not possess a
supermassive black hole. We note that the range of black hole masses in simulated red spirals found here is similar to
the one found by \citet{Davis2018} for the observed sample of a few tens of red spirals with directly measured
black hole masses. The trend of redder galaxies hosting more massive black holes is also present in observations
\citep{Dullo2020}. For the black hole accretion rates, the situation is the opposite: the red disks display much lower
accretion rates, with a median of $0.87 \times 10^{6}$ M$_\odot$ Gyr$^{-1}$ versus $3.6 \times 10^{6}$ M$_\odot$ Gyr$^{-1}$ for the remaining disks. The last histogram shows that the total AGN feedback energy
injected into the surrounding gas in the kinetic mode is typically much larger for the red disks than for the remaining
ones, with medians of $2.3 \times 10^{16}$ and $4.0 \times 10^{13}$ M$_\odot$ ckpc$^2$ Gyr$^{-2}$, respectively.
This means that the red disks typically grew their black holes earlier, reached the threshold of $10^{8}$ M$_\odot$,
and injected more of the feedback energy into the gas at earlier times, giving the galaxies enough time to become red.

Further justification for this line of reasoning can be provided by considering different characteristic scales of
galaxy evolution. The key parameter that controls the amount of star formation in the galaxy and thus its
transformation into a red object is the total gas mass and its evolution. As suggested by \citet{Xu2022}, the time
at which this mass reaches a maximum and starts to decrease, $t_{\rm gas}$, may be considered the onset of reddening. We
estimated the value of this parameter for all our red disks. Next, we considered other characteristic timescales,
namely the time when the black hole reaches the threshold of $10^{8}$ M$_\odot$, denoted as $t_{\rm BH}$, and the time
of maximum dark-matter mass, $t_{\rm DM}$, which, if different from the present time, signifies the onset of the
mass-stripping interaction.

Figure~\ref{cortimegasbh} illustrates the correlation between the characteristic timescales $t_{\rm BH}$ and $t_{\rm
gas}$ for 324 out of the 377 red disks with final black hole masses larger than $10^8$ M$_\odot$, that is,
those for which $t_{\rm BH}$ could be estimated. In addition, this subsample was divided into weakly and strongly
stripped galaxies, namely those that lost less and more than half of their maximum dark-matter masses (for example ID12
and ID296304 from Fig.~\ref{evolution}, respectively). Comparing the two panels we notice that there is a significant
correlation between $t_{\rm BH}$ and $t_{\rm gas}$ (with the Pearson correlation coefficient $r = 0.35$) for
the weakly stripped subsample (upper panel), while no correlation is seen for the strongly stripped galaxies (lower
panel, $r = -0.11$).

In Fig.~\ref{cortimegasdm} we plot in a similar way the timescales $t_{\rm DM}$ versus $t_{\rm gas}$ for all 377 red
disks. Again, the sample was divided into weakly and strongly stripped galaxies using the same criterion. Among the
weakly stripped sample in the upper panel, there are many galaxies that reach their maximum mass at present, and these
cluster along the line corresponding to $t_{\rm DM} = 13.8$ Gyr. In contrast to the results in the previous figure,
a much weaker correlation (with the correlation coefficient $r = 0.24$) is now seen in the upper panel for weakly
stripped galaxies, but there is a very significant correlation ($r = 0.71$) in the lower panel for the strongly stripped disks.
Interestingly, $t_{\rm gas}$ is usually lower than $t_{\rm DM}$ (most of the points lie above the diagonal line in both
panels), which may be at least partially due to the fact that gas is also depleted as a result of star formation.

\begin{figure*}
\centering
\includegraphics[width=6cm]{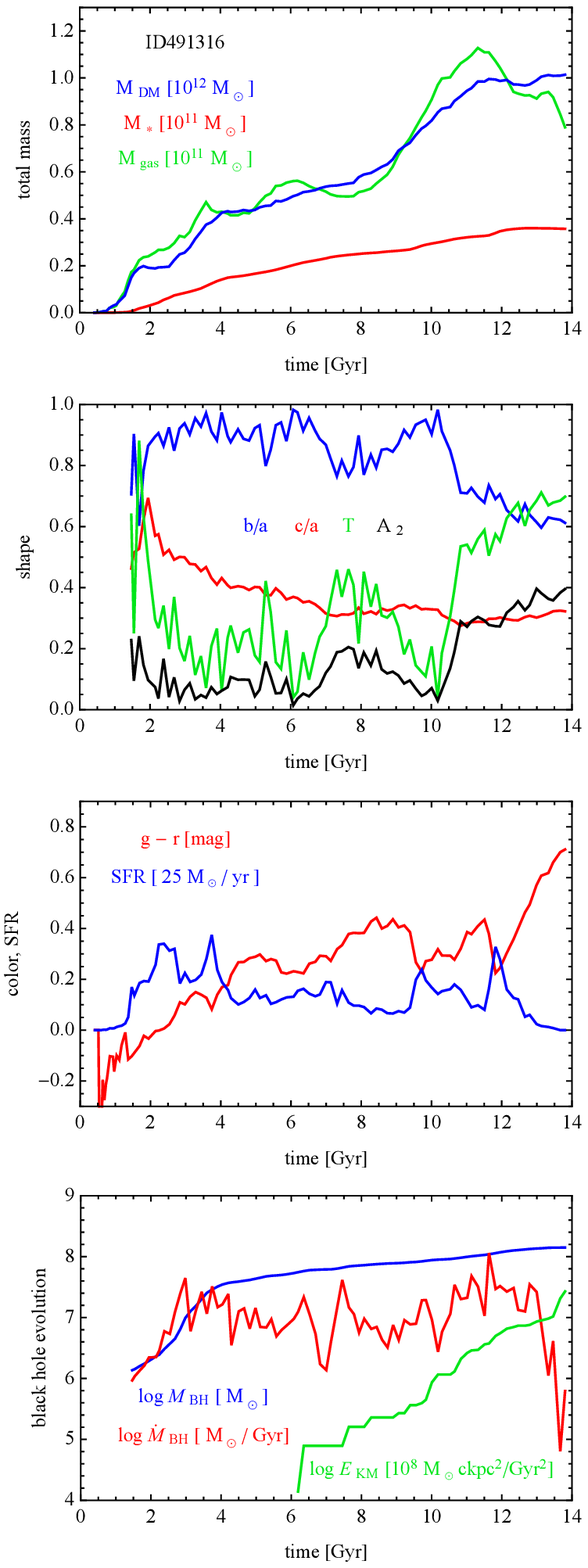}
\includegraphics[width=6cm]{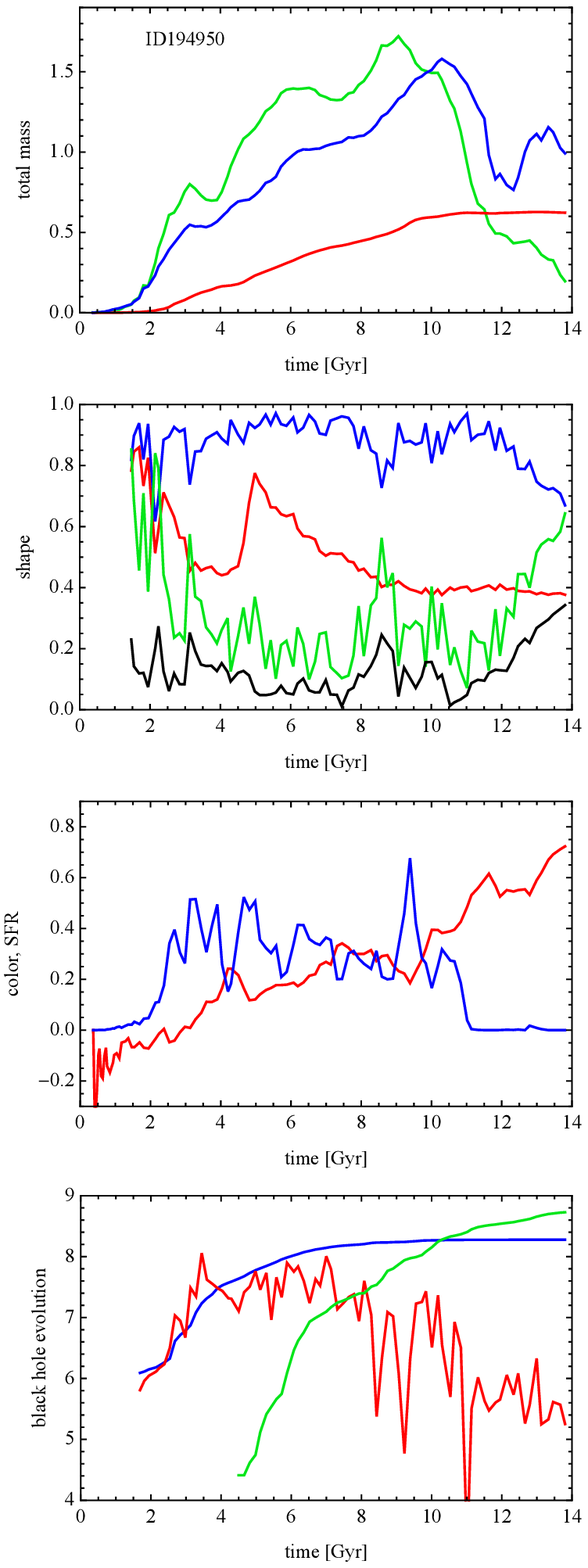}
\includegraphics[width=6cm]{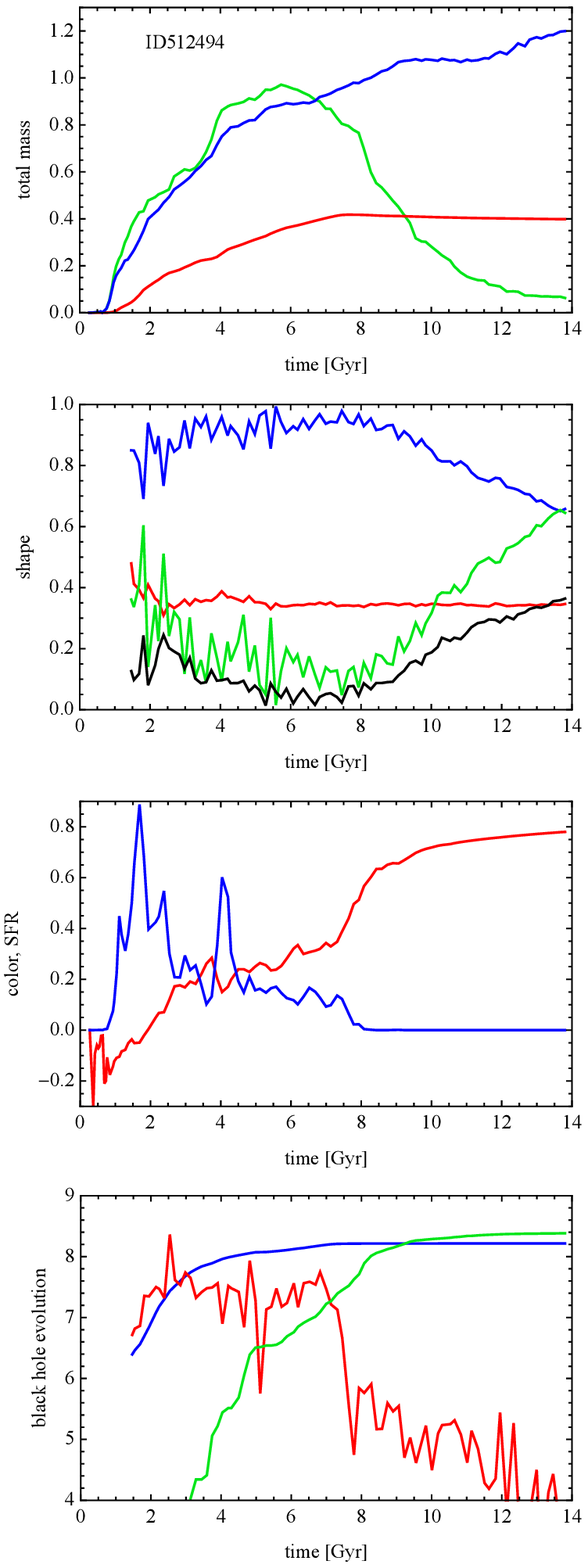}
\caption{Evolution of selected red disk galaxies with bars. The columns present the results for different
galaxies, ID491316, ID194950, and ID512494. First row: Evolution of the total dark,
stellar, and gas mass shown with the blue, red, and green lines, respectively. Second row: Evolution of four structural
properties of the galaxies: the axis ratios $b/a$ (blue line) and $c/a$ (red), the triaxiality parameter $T$
(green), and the bar strength $A_2$ (black). Third row: Evolution of color (red
line) and SFR (blue). Fourth row: Evolution of the mass of the black hole (blue), its accretion
rate (red), and the cumulative AGN feedback energy injected into
the surrounding gas in the kinetic mode (green). The units are given in the left column.}
\label{evolutionbars}
\end{figure*}

Putting together the results from the two figures, we can conclude that for isolated objects the loss of gas is most
often related to black hole feedback, while for the interacting ones the gas loss occurs at a similar time as the
dark-matter loss and is due to the ram pressure and tidal stripping. We note, however, that in addition to
the 53 galaxies with black hole masses below $10^8$ M$_\odot$, there is a significant number of isolated objects for
which $t_{\rm BH}$ and $t_{\rm gas}$ are very different (upper panel of Fig.~\ref{cortimegasbh}), and most of them have
$t_{\rm BH} > t_{\rm gas}$. In these galaxies the black holes grew slowly and the AGN feedback was probably not the
dominant factor in their quenching. These galaxies most likely became passive just by using up all the gas that was
available to them and stopping star formation.

\section{Red spirals with bars}

The analysis of the morphology of a large number of red spirals among the Galaxy Zoo objects revealed \citep{Masters2011}
that red spirals are more likely to possess bars than the general disk population. In this section we study the barred
galaxies from the sample of 377 red disks identified in the last output of the IllustrisTNG100 simulation.
Figure~\ref{surden} shows a few examples of strongly barred red disks selected in order to illustrate different
formation scenarios, which we discuss below. The plots show the face-on projections of the stellar surface density, with
bars clearly visible in the center. Interestingly, the bar in the galaxy ID491316 (left panel) is a little lopsided,
similar to the strongly lopsided bar-like galaxies studied by \citet{Lokas2021}.

The bars of the red disks show typical properties, including the profiles of the bar mode, $A_2$, shown in
Fig.~\ref{a2profiles}. This quantity is the $m=2$ mode of the Fourier decomposition of the surface density distribution
of stellar particles projected along the short axis, $A_m (R) = | \Sigma_j m_j \exp(i m \theta_j) |/\Sigma_j m_j$,
where $\theta_j$ is the azimuthal angle of the $j$th star, $m_j$ is its mass, and the sum goes up to the number of
particles in a given radial bin along the cylindrical radius, $R$. We can see that all three $A_2$ profiles of the
selected barred red disks are remarkably similar in terms of shape and maximum values. The decline in the $A_2$ profile
with radius can be used to estimate the length of the bar as the radius where the value of $A_2$ drops to half the
maximum. For all three galaxies the length of the bar is thus on the order of 6 kpc, in agreement with the visual
impression from the images shown in Fig.~\ref{surden}.

Figure~\ref{evolutionbars} illustrates different scenarios that lead to bar formation in red disks. The four rows
of the figure show the evolution of different parameters, such as the total mass (upper panels), the shape (second row
panels), color and SFR (third row), and the black hole mass, accretion rate, and the cumulative AGN feedback
energy injected into the surrounding gas in the kinetic mode (lower panels). The shape is described using the
shortest-to-longest and intermediate-to-longest axis ratios, $c/a$ and $b/a$, determined from the mass tensor of
stellar particles, as well as the triaxiality parameter $T = [1-(b/a)^2]/[1-(c/a)^2]$, all measured using stars
within $2 r_{1/2}$. We also plot the evolution of the global value of the bar mode, $A_2$, calculated using a single
radial bin, namely stars with radii $0 < R < 2 r_{1/2}$. We adopted the threshold of $A_2 > 0.2$ as corresponding to the
presence of the bar and defined the formation time of the bar as the time when $A_2$ crosses the value of 0.2
and remains above this threshold until the present time. The rest of the parameters are defined as in
Fig.~\ref{evolution}.

The left-column plots show an example of a galaxy (ID491316) in which bar formation was probably induced by
an interaction with a satellite of mass above $10^{10}$ M$_{\odot}$ that passes close to it on a prograde orbit. Such
orbits are known to lead to stronger morphological features than retrograde ones \citep{Lokas2018,
Peschken2019}. The first two (out of four) pericenter passages, around $t = 6$ and $t = 10$ Gyr at the distance of
$\sim30$ kpc, coincide with strong changes in the shape parameters. After the pericenters, the values of $T$ and $A_2$
increase, signifying the formation of a bar. While this change does not seem permanent after the first pericenter and
the bar is destroyed (probably due to gas accretion), after the second one it is much more pronounced, leading to the
formation of a strong and persistent bar.

During the interaction, the galaxy continues to possess a significant amount of gas, although mostly
in the outskirts. The gas mass starts to decrease only in the last few gigayears, which seems to be related to the black
hole reaching the threshold of $\sim10^8$ M$_\odot$. This causes the black hole to switch from thermal to kinetic mode
and blow the gas out from the center, after which the black hole accretion drops significantly (lower panel). The
removal of the gas quenches the galaxy, bringing its SFR down to zero and increasing its color above $g - r =0.6$. In
the case of ID491316, the bar formation precedes strong feedback from the black hole, and it could have funneled
the gas toward the center, feeding the black hole \citep{Emsellem2015}. The bar could also contribute to the quenching
of the galaxy by redistributing its gas in a process known as bar quenching \citep{Khoperskov2018}, which could
lead to the increase in the SFR around 12 Gyr.

In the galaxy ID194950 (middle column of Fig.~\ref{evolutionbars}), the bar forms even later, around $t = 12$ Gyr, and
coincides with strong gas and dark-matter mass loss related to an interaction with a bigger galaxy (also a red disk
included in the sample). There was also a merger that temporarily distorted the galaxy around $t = 8.5$ Gyr. In this
case, the black hole feedback was extended over a long period of time because the black hole reached the mass of
$\sim10^8$ M$_\odot$ early, around $t = 6$ Gyr. The quenching of star formation and the galaxy reddening coincide,
however, with the time of the interaction. It is thus difficult to say which of the effects was more important in
removing the gas and aiding bar formation.

\begin{figure}
\centering
\includegraphics[width=8cm]{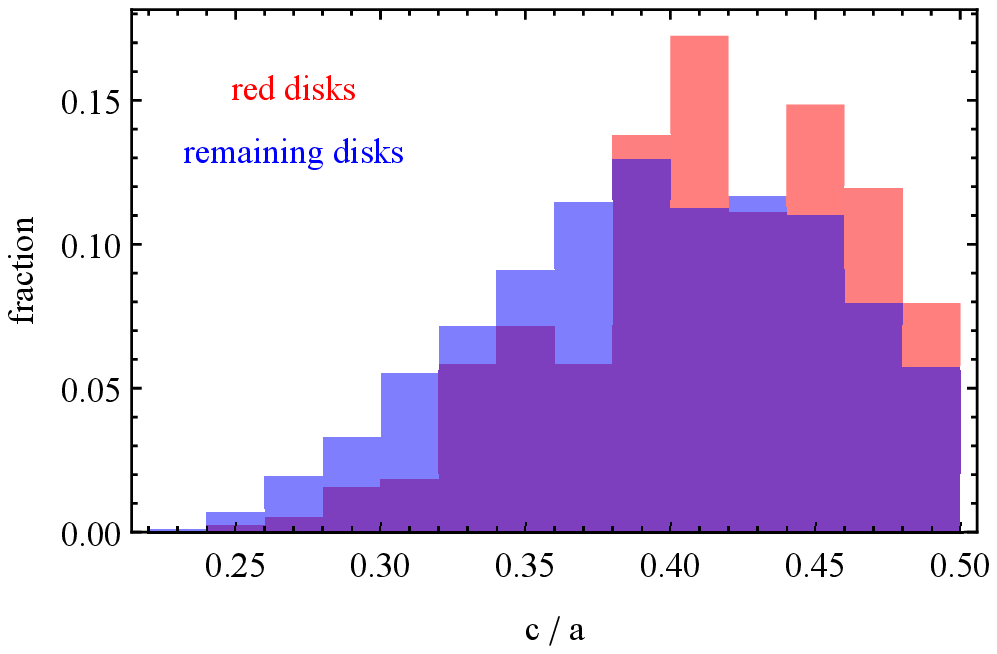}\\
\vspace{0.3cm}
\includegraphics[width=8cm]{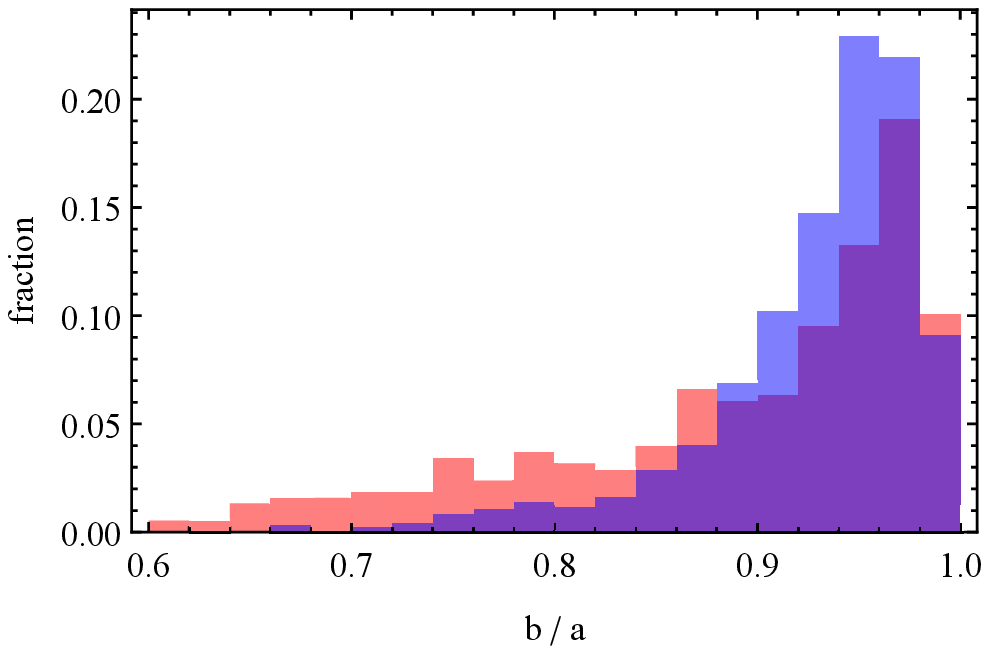}\\
\vspace{0.3cm}
\includegraphics[width=8cm]{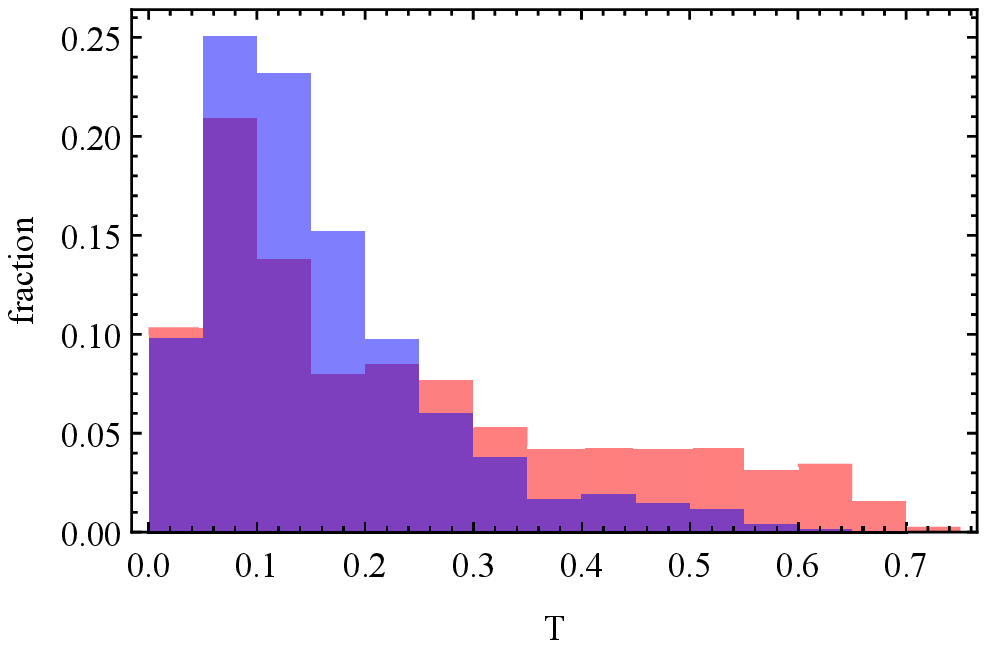}\\
\vspace{0.3cm}
\includegraphics[width=8cm]{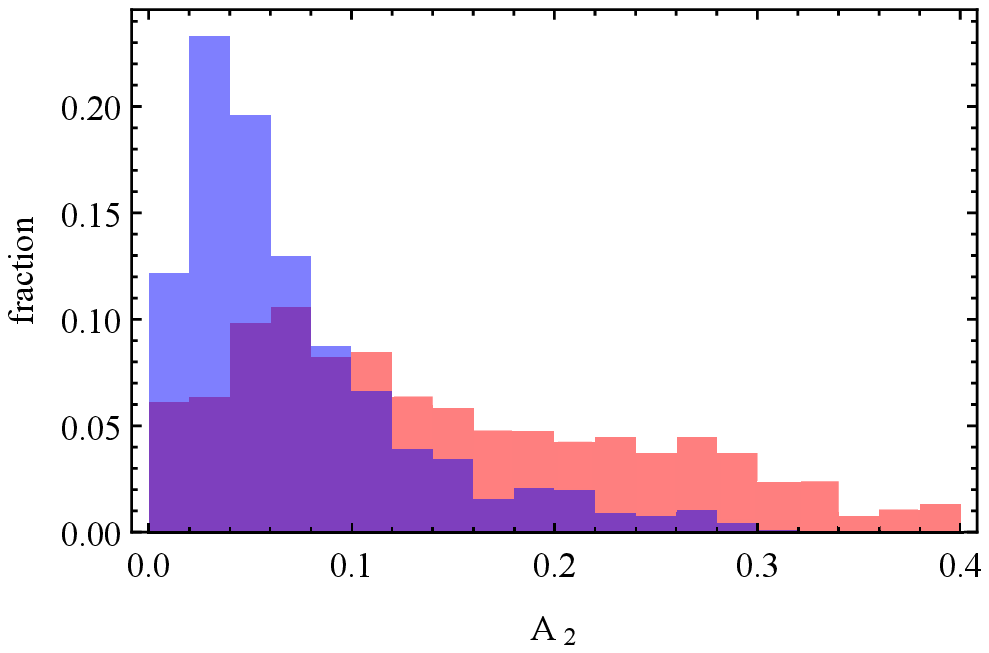}
\caption{Distributions of different shape parameters of disky galaxies. The four panels, from top to bottom, show
the distributions of the axis ratio $c/a$, the axis ratio $b/a$, the triaxiality parameter ($T$), and the bar mode ($A_2$)
for red disks with $g - r > 0.6$ (red) and remaining disks with $g - r < 0.6$ (blue). Measurements of the properties
were done within $2 r_{1/2}$, and the histograms were normalized to unity.}
\label{histogramsshape}
\end{figure}

\begin{figure}
\centering
\includegraphics[width=8cm]{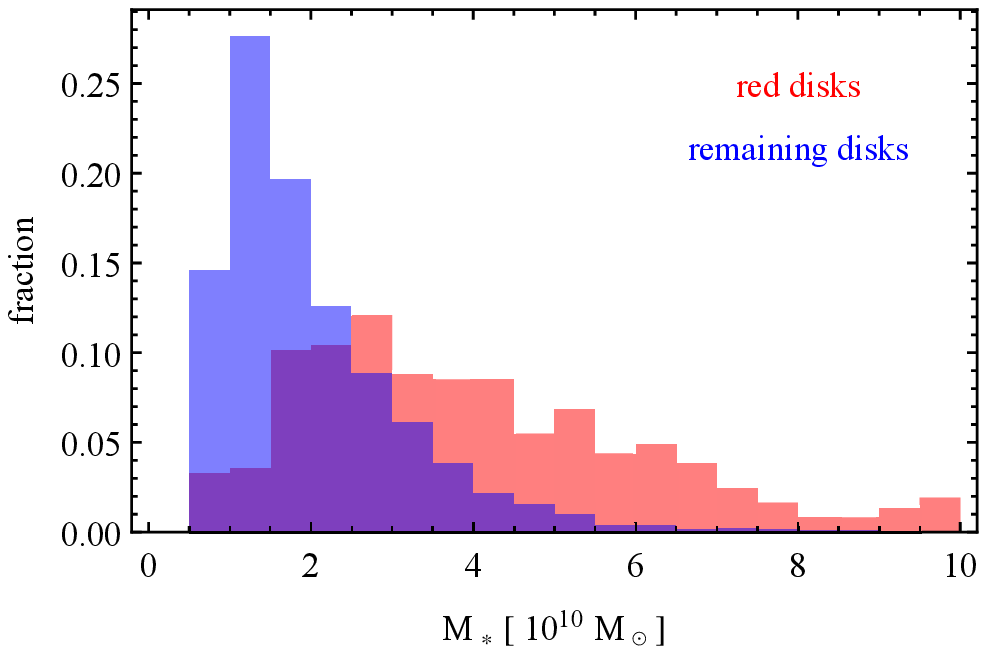}\\
\vspace{0.3cm}
\includegraphics[width=8cm]{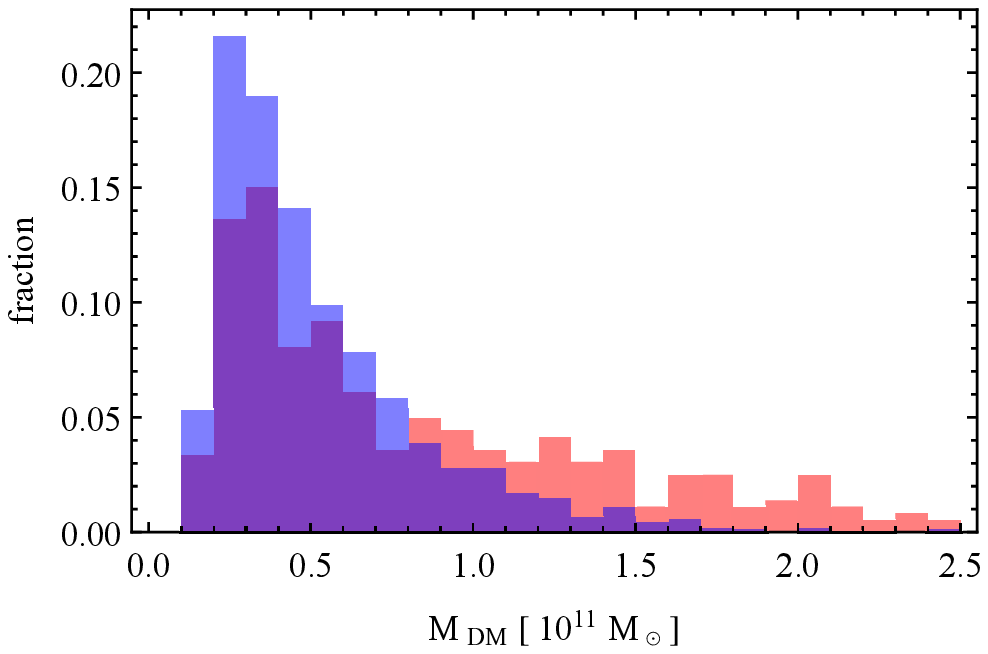}\\
\vspace{0.3cm}
\includegraphics[width=8cm]{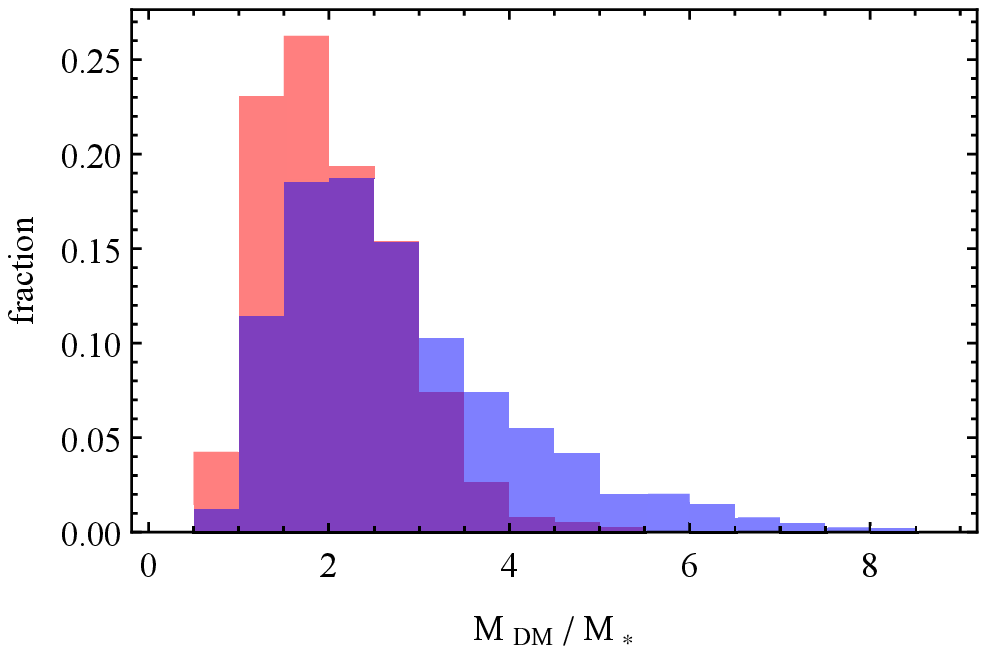}\\
\caption{Distributions of different mass measures of disky galaxies. The three panels, from top to bottom, show the
distributions of the stellar mass, the dark mass, and the ratio of the two for red disks with $g - r > 0.6$
(red) and remaining disks with $g - r < 0.6$ (blue). Measurements of the properties were done within $2 r_{1/2}$, and
the histograms were normalized to unity.}
\label{histogramsmass}
\end{figure}

The right column of Fig.~\ref{evolutionbars} illustrates yet another interesting scenario of bar formation in the red
disk ID512494. This galaxy had no significant interactions after a minor merger at $t = 6$ Gyr, which did not affect its
shape. Instead, around that time the feedback from the black hole seems to have started removing the gas from the
galaxy and, as a result, its inner part (within $2 r_{1/2}$) was completely devoid of gas by $t = 8$ Gyr.
Since there is no more fuel, the black hole accretion rate drops (lower panel); however, at exactly this time the bar
starts to grow (second panel). In this case, therefore, it seems that black hole feedback is the dominant factor behind
both the bar formation and the stopping of star formation (third panel). The total AGN feedback energy injected into
the surrounding gas in the kinetic mode is much higher in the cases of ID194950 and ID512494 than for ID491316 (lower
panels).

While it is difficult to ascertain which of these scenarios is the dominant way to form bars in red disks, they all
appear rather effective in causing the prevalence of bars in red spirals with respect to the rest of the disk
population. In Fig.~\ref{histogramsshape} we illustrate the differences in the shape of the members of the red disk
sample and the rest of the disks in terms of the axis ratios $c/a$ and $b/a$, the triaxiality parameter ($T$), and the bar
mode ($A_2$), all measured within $2 r_{1/2}$. We can see that the red disks are on average a little thicker
(larger $c/a$) than the rest and significantly more prolate (lower $b/a$ and higher $T$).

These differences are due to the fact that the red disks more often contain bars and that the bars are stronger, as shown by
the comparison of the distributions of the $A_2$ mode in the lower panel of the figure. We can see that the
distributions are significantly different, with median $A_2$ values of 0.12 for the red disks and 0.05 for the
remaining disks. Since we define barred galaxies as those with the global value $A_2 > 0.2$, we find 108 barred
galaxies among the 377 red disks and only 85 among the 1535 remaining disks, which corresponds to a bar fraction of
29\% and 6\%, respectively.

The efficiency of bar formation in galaxies is known to correlate with the stellar mass \citep{Masters2012} and the
stellar fraction in the inner part \citep{Athanassoula1986, Athanassoula2003}. In Fig.~\ref{histogramsmass} we
demonstrate that, indeed, the stellar masses of red disks within $2 r_{1/2}$, where the bars form, are on average
larger than for the remaining disks (upper panel). While the dark masses within this region are also slightly
larger (middle panel), the red disks are less dominated by dark matter than the rest of the disks (lower panel).

\section{Discussion}

We have studied the properties and origins of red galaxies selected from the sample of disky galaxies identified in the
final output of the IllustrisTNG100 simulation. The initial sample of 1912 disks included well-resolved galaxies with
sufficient numbers of stars to allow a proper morphological classification, which was performed on the basis of rotational
support and flat shape. Our final sample of red disks contained 377 objects selected with the sole criterion of color,
$g - r > 0.6$. The red disks identified in this fashion proved to be really passive, with very low gas fractions and
SFRs in comparison to the rest of the disks.

The fraction of red disks among all disk galaxies in the simulation is 20\%, which falls in between the values quoted
in the literature for the observed samples. In general, samples of red spirals are selected differently in particular
studies, which makes the comparison between the results difficult. Both the initial spiral galaxies and the red disks
among them are chosen using discrepant criteria. For example, our simple color cut of $g - r > 0.6$ was also applied by
\citet{Mahajan2020}, who found that red spirals comprise 42\% of disks in the GAMA survey, but a more stringent,
luminosity-dependent cut was used in \citet{Masters2010}, where the fraction of passive spirals was as low as 5\%.
These choices can partially explain the difference between the fractions estimated in different studies. The
much higher fraction of red spirals found by \citet{Mahajan2020} despite using the same color cut is probably due to
differences between the initial samples of disks. While the IllustrisTNG simulations provide a solid improvement over
the original Illustris results in terms of reproducing different galaxy populations and their properties, the agreement
is still not perfect and our simple method of selecting disk galaxies is certainly unable to adequately mimic the
procedures used by observers.

We also attempted to determine what mechanisms are responsible for the quenching of spiral galaxies in the simulation.
Perhaps surprisingly, we found that only about 13\% of the red disks experienced strong mass loss during their
evolution, having lost more than half of their maximum dark-matter mass. Such mass loss is indicative of a strong
interaction with a more massive structure that can cause gas stripping, either by ram pressure or its
combination with tidal effects. We emphasize that it is difficult to disentangle the two processes because
they both occur in dense environments. In particular, it would be hard to imagine a configuration where only ram
pressure stripping would operate since high gas densities are usually associated with massive objects. For the
remaining red disks, AGN feedback seems to play a dominant role, which is supported by the higher black hole masses,
lower accretion rates, and larger AGN feedback energies of red disks in comparison to the
rest. The characteristic timescales of gas loss also correlate more convincingly with the timescales of black hole
growth rather than those of mass loss.

There seems to be a consensus in the literature by now that no single mechanism can be identified as the unique cause of
the quenching of spiral galaxies. The two mechanisms we discussed were recently agreed upon as the most promising:
red disks are believed to have formed via AGN feedback and/or via environmental effects in clusters, but the jury is
still out on the question of which of these mechanisms is the dominant one. Contrary to our results, \citet{Hughes2009}
claim that environment is more important than AGN feedback for their observed sample, while \citet{Fraser2018} find
evidence for both quenching via internal processes and environment in their passive spirals. \citet{Shimakawa2022}
find the average passive fraction of spiral galaxies in clusters to be approximately three times higher than in random
fields. In particular, they detected a substantial excess of passive spiral galaxies at intermediate distances from the
cluster center. Still, their results suggest that environmental effects cannot be the only ones responsible for
creating red spirals, though they can help keep them quiescent.

The prevalence of AGN feedback as the main quenching mechanism in red spirals that we found here is in accordance with
\citet{Weinberger2018}, who demonstrated that the quenching of all massive central galaxies in IllustrisTNG coincides
with the onset of the kinetic mode feedback. Similar conclusions were reached by \citet{Xu2022} for the subsample of
massive isolated red spirals from TNG300. Here we extended these results by considering not only isolated red spirals
but the whole population; we benefited from the higher resolution of the TNG100 run, which allowed for a more reliable
selection of disk galaxies and a detailed look into morphological features such as bars. While \citet{Xu2022} used
binned samples of galaxies quenched at different epochs to demonstrate the correlation between the decrease in the
black hole accretion rate and the quenching time, here we used a more direct approach and find a correlation between
the quenching time and the timescale of black hole growth measured for each galaxy. Our conclusions, however, are
subject to the caveat related to the over-quenching effect known to exist in the IllustrisTNG simulations
\citep{Angthopo2021} and to the reliability of the model for black hole feedback in general. An ultimate test of this
model is not yet possible, but it was demonstrated that it leads to a better agreement of galaxy properties with
observations in comparison with the original spherical bubble model of Illustris \citep{Weinberger2017, Weinberger2018}.

The red spiral galaxies we identified in the IllustrisTNG100 simulation are much more likely to possess bars than the
remaining disks, and their bars are typically stronger. This finding is in very good agreement with the results from the
Galaxy Zoo survey \citep{Masters2010, Masters2011, Masters2012} despite the fact that the overabundance of bars
among red and gas-poor spirals was questioned and interpreted as an observational bias by \citet{Erwin2018}. We note,
however, that our simulated sample and those of the Galaxy Zoo are restricted to rather massive spiral galaxies.

The causal relation between the formation of bars and quenching in red spirals remains unclear. Among our barred red
disks, we identified cases of bars forming prior to, simultaneously with, and after quenching. The mechanism of
quenching induced by bars was invoked as one of the possible mechanisms for the formation of red spirals. Although the
details of gas evolution in the bar region cannot be sufficiently resolved in the IllustrisTNG100 simulations, it seems
unlikely that this mechanism is operating in our red disks. According to the implementation of this mechanism proposed
by \citet{Khoperskov2018}, bars induce turbulence in the gas and stabilize it against collapse, increasing its
dispersion rather than expelling it out of the galaxy. Instead, in the simulated red disks studied here, the gas is
typically removed from the central part of the galaxy, which is consistent with the effect of AGN feedback. The
correlation between the presence of the bar and quenching may thus instead originate from the fact that bars are more
likely to form in galaxies with low gas content \citep{Athanassoula2013, Lokas2020a}.

\begin{acknowledgements}
I am grateful to the anonymous referee for useful comments and to the IllustrisTNG team for making their
simulations publicly available.
\end{acknowledgements}


\begin{thebibliography}{}

\bibitem[{Angthopo et al.}(2021)]{Angthopo2021} Angthopo, J., Negri, A., Ferreras, I., et al. 2021, MNRAS, 502, 3685
\bibitem[{Athanassoula}(2003)]{Athanassoula2003} Athanassoula, E. 2003, MNRAS, 341, 1179
\bibitem[{Athanassoula \& Sellwood}(1986)]{Athanassoula1986} Athanassoula, E., \& Sellwood, J. A. 1986, MNRAS, 221, 213
\bibitem[{Athanassoula et al.}(2013)]{Athanassoula2013} Athanassoula, E., Machado, R. E. G., \& Rodionov, S. A. 2013,
        MNRAS, 429, 1949
\bibitem[{Bekki et al.}(2002)]{Bekki2002} Bekki, K., Couch, W. J., Shioya, Y. 2002, ApJ, 577, 651
\bibitem[{Bonne et al.}(2015)]{Bonne2015} Bonne, N. J., Brown, M. J. I., Jones, H., \& Pimbblet, K. A.
        2015, ApJ, 799, 160
\bibitem[{Cortese}(2012)]{Cortese2012} Cortese, L. 2012, A\&A, 543, A132
\bibitem[{Couch et al.}(1998)]{Couch1998} Couch, W. J., Barger, A. J., Smail, I., Ellis, R. S., \& Sharples, R. M.
        1998, ApJ, 497, 188
\bibitem[{Davis et al.}(2018)]{Davis2018} Davis, B. L., Graham, A. W., \& Cameron, E. 2018, ApJ, 869, 113
\bibitem[{Dullo et al.}(2020)]{Dullo2020} Dullo, B. T., Bouquin, A. Y. K., Gil de Paz, A., Knapen, J. H., \& Gorgas, J.
        2020, ApJ, 898, 83
\bibitem[{Emsellem et al.}(2015)]{Emsellem2015} Emsellem, E., Renaud, F., Bournaud, F., et al. 2015, MNRAS, 446, 2468
\bibitem[{Erwin}(2018)]{Erwin2018} Erwin, P. 2018, MNRAS, 474, 5372
\bibitem[{Fraser-McKelvie et al.}(2018)]{Fraser2018} Fraser-McKelvie, A., Brown, M. J. I., Pimbblet, K., Dolley, T., \&
        Bonne, N. J. 2018, MNRAS, 474, 1909
\bibitem[{Genel et al.}(2015)]{Genel2015} Genel, S., Fall, S. M., Hernquist, L., et al. 2015, ApJ, 804, L40
\bibitem[{Genel et al.}(2018)]{Genel2018} Genel, S., Nelson, D., Pillepich, A., et al. 2018, MNRAS, 474, 3976
\bibitem[{Goto et al.}(2003)]{Goto2003} Goto, T., Okamura, S., Sekiguchi, M., et al. 2003, PASJ, 55, 757
\bibitem[{Gunn \& Gott}(1972)]{Gunn1972} Gunn, J. E., \& Gott, J. R. 1972, ApJ, 176, 1
\bibitem[{Haslbauer et al.}(2022)]{Haslbauer2022} Haslbauer, M., Banik, I., Kroupa, P., Wittenburg, N., \&
        Javanmardi, B. 2022, ApJ, 925, 183
\bibitem[{Hubble}(1936)]{Hubble1936} Hubble, E. P. 1936, The Realm of the Nebulae (Yale Univ. Press, New Haven)
\bibitem[{Hughes \& Cortese}(2009)]{Hughes2009} Hughes, T. M., \& Cortese, L. 2009, MNRAS, 396, L41
\bibitem[{Joshi et al.}(2020)]{Joshi2020} Joshi, G. D., Pillepich, A., Nelson, D., et al. 2020, MNRAS, 496, 2673
\bibitem[{Khoperskov et al.}(2018)]{Khoperskov2018} Khoperskov, S., Haywood, M., Di Matteo, P., Lehnert, M. D., \&
        Combes, F. 2018, A\&A, 609, A60
\bibitem[{{\L}okas}(2018)]{Lokas2018} {\L}okas, E. L. 2018, ApJ, 857, 6
\bibitem[{{\L}okas}(2020a)]{Lokas2020a} {\L}okas, E. L. 2020a, A\&A, 634, A122
\bibitem[{{\L}okas}(2020b)]{Lokas2020b} {\L}okas, E. L. 2020b, A\&A, 638, A133
\bibitem[{{\L}okas}(2021)]{Lokas2021} {\L}okas, E. L. 2021, A\&A, 655, A97
\bibitem[{{\L}okas}(2022)]{Lokas2022} {\L}okas, E. L. 2022, A\&A, 662, A53
\bibitem[{Mahajan et al.}(2020)]{Mahajan2020} Mahajan, S., Gupta, K. K., Rana, R., et al. 2020, MNRAS, 491, 398
\bibitem[{Marinacci et al.}(2018)]{Marinacci2018} Marinacci, F., Vogelsberger, M., Pakmor, R., et al. 2018,
        MNRAS, 480, 5113
\bibitem[{Masters et al.}(2010)]{Masters2010} Masters, K. L., Mosleh, M., Romer, A. K., et al. 2010, MNRAS, 405, 783
\bibitem[{Masters et al.}(2011)]{Masters2011} Masters, K. L., Nichol, R. C., Hoyle, B., et al. 2011, MNRAS, 411, 2026
\bibitem[{Masters et al.}(2012)]{Masters2012} Masters, K. L., Nichol, R. C., Haynes, M. P., et al.
        2012, MNRAS, 424, 2180
\bibitem[{Naiman et al.}(2018)]{Naiman2018} Naiman, J. P., Pillepich, A., Springel, V., et al., 2018, MNRAS, 477, 1206
\bibitem[{Nelson et al.}(2018)]{Nelson2018} Nelson, D., Pillepich, A., Springel, V., et al. 2018, MNRAS, 475, 624
\bibitem[{Nelson et al.}(2019)]{Nelson2019} Nelson, D., Springel, V., Pillepich, A., et al. 2019,
        Computat. Astroph. Cosmol., 6, 2
\bibitem[{Okamoto et al.}(2008)]{Okamoto2008} Okamoto, T., Nemmen, R. S., \& Bower, R. G. 2008, MNRAS, 385, 161
\bibitem[{Peng et al.}(2015)]{Peng2015} Peng, Y., Maiolino, R., \& Cochrane, R. 2015, Nature, 521, 192
\bibitem[{Peschken \& {\L}okas}(2019)]{Peschken2019} Peschken, N., \& {\L}okas, E. L. 2019, MNRAS, 483, 2721
\bibitem[{Pillepich et al.}(2018)]{Pillepich2018} Pillepich, A., Nelson, D., Hernquist, L., et al. 2018,
        MNRAS, 475, 648
\bibitem[{Poggianti et al.}(1999)]{Poggianti1999} Poggianti, B. M., Smail, I., Dressler, A., et al. 1999, ApJ, 518, 576
\bibitem[{Rodriguez-Gomez et al.}(2019)]{Rodriguez2019} Rodriguez-Gomez, V.,  Snyder, G. F., Lotz, J. M., et al.
        2019, MNRAS, 483, 4140
\bibitem[{Rowlands et al.}(2012)]{Rowlands2012} Rowlands, K., Dunne, L., Maddox, S., et al. 2012, MNRAS, 419, 2545
\bibitem[{Shimakawa et al.}(2022)]{Shimakawa2022} Shimakawa, R., Tanaka, M., Bottrell, C., et al. 2022, PASJ, 74, 612
\bibitem[{Springel et al.}(2001)]{Springel2001} Springel, V., White, S. D. M., Tormen, G., \& Kauffmann, G.
        2001, MNRAS, 328, 726
\bibitem[{Springel et al.}(2018)]{Springel2018} Springel, V., Pakmor, R., Pillepich, A., et al. 2018, MNRAS, 475, 676
\bibitem[{van den Bergh}(1976)]{Bergh1976} van den Bergh, S. 1976, ApJ, 206, 883
\bibitem[{Weinberger et al.}(2017)]{Weinberger2017} Weinberger, R., Springel, V., Hernquist, L., et al.
        2017, MNRAS, 465, 3291
\bibitem[{Weinberger et al.}(2018)]{Weinberger2018} Weinberger, R., Springel, V., Pakmor, R., et al.
        2018, MNRAS, 479, 4056
\bibitem[{Wolf et al.}(2009)]{Wolf2009} Wolf, C., Aragon-Salamanca, A., Balogh, M., et al. 2009, MNRAS, 393, 1302
\bibitem[{Xu et al.}(2022)]{Xu2022} Xu, Y., Luo, Y., Kang, X., et al. 2022, ApJ, 928, 100


\end{thebibliography}
\end{document}